\documentclass[11pt]{article}

\usepackage{natbib}

\usepackage{amsmath}
\usepackage{amssymb}
\usepackage{amsthm}
\usepackage{bbm}
\usepackage{bm}
\usepackage{graphicx}
\usepackage{apalike}
\usepackage{booktabs}
\usepackage{caption}
\usepackage{subcaption}
\usepackage{a4wide}
\usepackage{multirow}
\usepackage{booktabs}
\usepackage{siunitx}

\theoremstyle{definition}

\newtheorem{assumption}{Assumption}

\newcommand{\E}{\mathbb E}
\newcommand{\e}{\mathrm e}

\newcommand{\B}{\bm{\mathrm{B}}}

\newcommand{\M}{\bm{\mathrm{M}}}

\newcommand{\Z}{\bm{\mathrm{Z}}}

\newcommand{\Dg}{\mathrm{Dg}}

\newcommand{\D}{\mathrm{d}}

\begin{document}
	
%\begin{frontmatter}
\title{Recurrent neural network based parameter estimation of Hawkes model on high-frequency financial data}
	
\author{Kyungsub Lee\footnote{Department of Statistics, Yeungnam University, Gyeongsan, Gyeongbuk 38541, Korea}}
\maketitle
\begin{abstract}

This study examines the use of a recurrent neural network for estimating the parameters of a Hawkes model based on high-frequency financial data, and subsequently, for computing volatility. 
Neural networks have shown promising results in various fields, and interest in finance is also growing.
Our approach demonstrates significantly faster computational performance compared to traditional maximum likelihood estimation methods while yielding comparable accuracy in both simulation and empirical studies. 
Furthermore, we demonstrate the application of this method for real-time volatility measurement, 
enabling the continuous estimation of financial volatility as new price data keeps coming from the market.

\end{abstract}

%\begin{keyword}
%    Hawkes model \sep neural network \sep estimation \sep volatility
%	\JEL  C51 \sep C58 
%\end{keyword}

%\end{frontmatter}

\section{Introduction}

We propose a real-time estimation and volatility measurement scheme.
Specifically, we continuously estimate the financial model parameters and volatility in real-time 
as new price process become available during the market operation.
Our method uses a recurrent neural network to estimate the parameters of a Hawkes model based on high-frequency financial data, 
and these estimates are subsequently used to compute volatility.
This approach exhibits significantly faster computational performance compared to the traditional maximum likelihood estimation (MLE) method.

The Hawkes process introduced by \cite{Hawkes1} is a type of point process used to model the occurrence of events over time.
It is frequently utilized in finance to model the arrival of trades or other financial events in the market; 
here, we use it to describe the fluctuations in the price in the tick structure. 
As a two dimensional Hawkes process with symmetric kernel in our study, 
it is characterized by two key features: self-excitation and mutual-excitation \citep{Bacryetal2013}. 
Self-excitation means that the occurrence of an event increases the likelihood of the same types of events occurring, 
while mutual-excitation implies that the occurrence of one event can increase the likelihood occurrence of other types of events. 

For the estimation, we use a long short-term memory (LSTM) network introduced by \cite{hochreiter1997long}.
This is a type of recurrent neural network that is capable of learning long-term dependencies in data.
LSTMs can be used for a variety of tasks that involve sequential data, including language modeling, machine translation, speech recognition, time series based on historical data, and financial analysis \citep{zhang2021role, ghosh2022forecasting}.
They are particularly useful for tasks where capturing long-term dependencies is important, 
as the gating mechanism allows the network to retain important earlier information in the sequence 
while discarding irrelevant or outdated information.

Thus, our method uses a neural network for parameter estimation.
Similar attempts have been made in recent years in various fields \citep{wlas2008neural, WANG2022128146, wei2022estimating}, 
but there is still limited research on using neural networks for estimation in financial time series.
Specifically, we use a direct parameter estimation approach in which the neural network is trained to directly predict the model parameters based on the observed data.
This method is an example of a small aspect of network-based parameter estimation and expected to be applied to more complex models.

%The remainder of the paper is organized as follows:
%Section~\ref{Sect:model} explains the financial price dynamics and network models.
%Section~\ref{Sect:simul} and \ref{Sect:empirical} presents the results of the simulation and empirical studies, respectively.
%Section~\ref{Sect:concl} presents the conclusions of this study.

\section{Method}~\label{Sect:model}

\subsection{Price model}

First, we explain the stochastic model to describe high-frequency stock price movements.
We model the up and down movements of the tick-level price process as a marked Hawkes process, 
which captures both the timing and size of the movements. 
This process is defined by the random measures,
\begin{equation}
\bm{M}(\D u \times \D z) = \begin{bmatrix} M_1(\D u \times \D z_1) \\ M_2(\D u \times \D z_2) \end{bmatrix} \label{Eq:measure}
\end{equation}
in the product space of time and jump size, $\mathbb{R} \times E_i$, for $i=1, 2$,
where $E_i = \mathbb{N} \times \{i\}$ denotes the space of mark (jump) sizes for up and down price movements, respectively. 
Each measure in Eq.~\eqref{Eq:measure} is associated with a sequence of $E_i$-valued random variables $\{Z_{i,n}\}$ in addition to the sequence of random times $\{\tau_{i, n}\}$ for each $i$.
That is,
$$M_i(\D u \times \D z_i) = \sum_{n} \delta_{\tau_{i,n}, Z_{i,n}}(\D u \times \D z_i)$$
with the Dirac measure $\delta$, which is defined as follows:
for any time interval $I$ and $A_i \subset E_i$
$$ \delta_{\tau_{i,n}, Z_{i,n}} (I \times A_i) =
\left\{\begin{array}{lr}
	1, \text{ if } \tau_{i,n} \in I \text{ and } Z_{i,n} \in A_i,  \\
	0, \text{ otherwise.}
\end{array}\right.
$$

A vector of c\`adl\`ag counting processes is defined by
$$ \bm{N}_t = \begin{bmatrix}
	N_1(t) \\
	N_2(t)
\end{bmatrix} = \int_{(0,t] \times E} \Dg(z) \bm{M}(\D u \times \D z), \quad \Dg(z) = \begin{bmatrix} z_1 & 0 \\ 0 & z_2 \end{bmatrix}, \quad E = E_1 \cup E_2$$
which counts the number of events weighted by their size, that is,
$$N_i(t) = N_i((0,t]) = \sum_{n} Z_{i,n} \mathbbm{1}_{ \{ 0 < \tau_{i,n} \leq t \} } = \# \textrm{ of } \tau_{i,n} \in (0, t], \quad \textrm{for } i=1,2.$$

\begin{assumption}
The stochastic intensity $\lambda_i$ for $N_i$ is represented by the following:
\begin{equation}
	\bm{\lambda}_t =
	\begin{bmatrix}
		\lambda_1(t) \\
		\lambda_2(t)
	\end{bmatrix} = 
	\bm{\mu}
	+ \int_{(-\infty,t] \times E}
	\bm{\alpha}
	\circ \bm{b} (t - u)
	  \bm{M}(\D u \times \D z) \label{Eq:lambda}
\end{equation}
where $\bm{\mu}$ is a $2 \times 1$ positive constant base intensity vector,
$\bm{\alpha}$ is a positive $2\times 2$ constant matrix.
$\bm{h}$ is a decay function matrix and $\circ$ denotes the element-wise product.
From the definition of Eq.~\eqref{Eq:lambda}, 
for simplicity, we assume that the future impact of an event on intensities is independent of jump size,
as the integrand of the equation does not contain the jump variable $z$.
In addition, for further parsimony, we assume that:
\begin{equation}
\bm{\mu} = \begin{bmatrix} \mu \\ \mu \end{bmatrix}, \quad
\bm{\alpha} = \begin{bmatrix} \alpha_1 & \alpha_2 \\ \alpha_2 & \alpha_1 \end{bmatrix}, \quad 
\bm{b}(t) = \begin{bmatrix}
	\e^{-\beta t} & \e^{-\beta t} \\
	\e^{-\beta t} & \e^{-\beta t}
\end{bmatrix}.
\end{equation}
Hence, the set of parameters to be estimated is $\{\mu, \alpha_1, \alpha_2, \beta \}$.
We also assume that the mark $Z_i$ at time $t$ is independent from the $\sigma$-algebra generated by $(N_j(s), \lambda_j(s))_{s < t}$ for $j = 1, 2$.
The intensity process is assumed to be stationary and that the spectral radius of $| \int_0^\infty \bm{\alpha} \circ \bm{b} (t) \D t|$ is less than 1.
\end{assumption}

Under this assumption, the Hawkes volatility of price movements -- the standard deviation of total up and down net movements -- is represented by
\begin{equation}
\mathrm{SD}(N_1(t) - N_2(t)) =  \sqrt{\bm{\mathrm{u}}^{\top} \left[ \mathcal{T} \left\{\overline \Z \circ \B  \right\}  +  \overline \Z^{(2)}\circ \Dg (\E [\bm{\lambda}_t]) \right] \bm{\mathrm{u}} t }, \quad \bm{\mathrm{u}} = \begin{bmatrix} 1 \\ -1 \end{bmatrix} \label{Eq:Var}
\end{equation}
where $\mathcal{T}$ is an operator such that $ \mathcal{T}(\M) = \M + \M^{\top}$ 
for a square matrix $\M$ and
$\mathrm{Dg}(\cdot)$ denotes a diagonal matrix whose diagonal entry is composed of the argument.
Furthermore,
\begin{equation}
	\E[\bm{\lambda}_t] = ( \bm{\beta} - \bm{\alpha} )^{-1}\bm{\beta}\bm{\mu},  \quad 
	\bm{\beta} = \begin{bmatrix} \beta & 0 \\ 0 & \beta  \end{bmatrix} \label{Eq:E_lambda2}
\end{equation}
and
\begin{equation}
\E[\bm{\lambda}_t\bm{\lambda}_t^{\top}] = (\bm{\beta} - \bm{\alpha})^{-1} \left( \frac{1}{2}\bm{\alpha} \Dg(\E[\bm{\lambda}_t]) \bm{\alpha} + \bm{\beta}\bm{\mu}\E[\bm{\lambda}_t^{\top}] \right) 
\end{equation}
and
\begin{equation}
	\B =  \left\{ \overline \Z^{\top}  \circ \E[\bm{\lambda}_t\bm{\lambda}_t^{\top}] + \Dg(\E[\bm{\lambda}_t]) \left(\bm{\alpha} \circ \overline \Z \right)^{\top} - \Dg(\overline \Z) \E[\bm{\lambda}_t] \E[\bm{\lambda}_t]^{\top} \right\} (\bm{\beta} - \bm{\alpha})^{-1}
\end{equation}
and by the mark independent assumption,
$$ 
\overline \Z = \begin{bmatrix} \E[Z_1] & \E[Z_2]  \\ \E[Z_1] & \E[Z_2]  \end{bmatrix},
\quad 
\overline \Z = \begin{bmatrix} \E[Z_1^2] & \E[Z_2^2]  \\ \E[Z_1^2] & \E[Z_2^2]  \end{bmatrix}.
$$
To calculate the volatility of price changes, rather than the number of movements, 
we multiply the minimum tick size to Eq.~\eqref{Eq:Var}.
Further details can be found in \cite{lee2017marked} and \cite{LeeHawkesVol}.

\subsection{Network model}

Next, we construct a recurrent neural network for parameter estimation.
The traditional method of estimating the parameters of a Hawkes process is MLE, 
which involves maximizing the log-likelihood function of the model:
$$
L(T, \bm\theta) = 	\sum_{i=1}^2 \left( \int_{(0, T] \times E} \log \lambda_{i}(u)  M_{i}(\D u \times \D z_i)  - \int_0^T \lambda_{i}(u)  \D u \right)
$$
to estimate the parameters most likely to generate the observed data.

In contrast, neural network based parameter estimation involves training a neural network to predict the parameters of a Hawkes process based on input data. 
To do this, numerous sample paths of inter-arrival times and movement types (up or down) are generated, 
where the parameters of each path are determined randomly. 
These sample paths are then used as feature variables, and the associated true parameter values are used as the target variables.

The neural network is trained on these data.
Once trained, 
it can be used to predict the parameter values of a new sample path of Hawkes process data. 
These predicted parameter values can then be used to compute further complicated formulae, such as the Hawkes volatility in Eq.~\eqref{Eq:Var}, 
which is a measure of the variability of the process over time.

Here, we use an LSTM model with three layers as the neural network. 
The LSTM network is known for its capability of retaining information over a long duration, 
making it appropriate for tasks that require context comprehension or state preservation. 
The gates mechanism in the LSTM architecture that regulates the flow of information between memory cells and the network
enables the network to choose which information should be preserved or discarded. 
We also tested gated recurrent unit networks \citep{cho2014properties};
however, the LSTM performed slightly better in our problem.
A thorough account of the network's implementation is presented in the following section.

\section{Simulation result}~\label{Sect:simul}

In this simulation study, we generate a set of paths of Hawkes processes to create a training dataset for the neural network. 
The dataset comprises a sufficient quantity of synthetic data, 
which are utilized to train the network to predict the real parameters of the Hawkes process. 
The real parameters for each Hawkes process are randomly selected to cover the entire range of possible values, with each path having distinct parameters. 
For the ranges of the parameters,
we use the ranges of the estimates obtained by fitting the past intraday price process of various stocks to the Hawkes model.
Approximately 30 symbols of stocks, including AAPL, AMZN, C, FB, GOOG, IBM, MCD, MSFT, NVDA, and XOM from 2018 to 2019 are used.

For the estimation, we focus on high-frequency rather than ultra-high-frequency.
More precisely, the raw data are filtered as follows.
We observe the mid-price at intervals of $\Delta t = 0.1$ seconds, noting any changes from the previously observed price. 
If a change is detected, we record the exact time of the change and the new price. 
If the price remains the same, we move on to the next interval of 0.1 seconds and repeat the process. 
This method allows us to filter out unnecessary movements, commonly referred to as microstructure noise, observed at ultra-high frequencies.

Once the set of estimates is obtained, we generate Hawkes process paths of a 2,000-time step. 
These are then used for neural network training together with estimates as target variables.
This method yields tens of thousands of datasets, which are sufficient to construct a comprehensive training set.

The implementation of the LSTM network model is as follows.
The first layer consists of 12 units, which manage sequential data.
These data are a two-dimensional input of time series data that comprise inter-arrivals and types of movements (up or down).
Up and down movements are encoded as 1 and 2, respectively.
The output of this layer is a sequence of 12 length of vectors, 
where each vector is the output of the first layer at a given time step.
Thus, if the original time series has a 2,000-time step, then the output of the first layer is a $2,000 \times 12$ matrix,
which is the time step $\times$ the number of units.

The second layer has 12 units and produces a single (not a sequence of) vector of length 12. 
The final layer is a dense (fully connected) layer with four units,
which produces output representing the parameters in the Hawkes model, $\mu, \alpha_1, \alpha_2$ and $\beta$.
If we extend the model for more complexity, 
the number of units in the last dense layer will be adjusted accordingly.
This is because each unit in the last layer represents each parameter.

As pointed out by \cite{mei2017neural}, a natural extension of LSTM is deep LSTM, 
especially for complex structured data such as high-frequency financial data, 
where the effectiveness of multi-layering seems promising. 
However, we utilized a relatively parsimonious Hawkes model and achieved sufficient performance without employing a large number of layers, 
so we used the LSTM model proposed above, which is relatively faster to train. 
If the data structure and model become more complex, an extension to deep LSTM would be helpful.

For training, we use 75,000 sample data points;
hence, the dataset for the neural network's input has a shape of $75,000 \times 1,000 \times 2$.
The Adam \citep{kingma2014adam} optimizer is used for training.
Generally, more than 300 epochs are used.
For testing, we use 15,000 data points that were not used for training.
This is done to evaluate the model's ability to make predictions on these unseen data based on mean squared error (MSE).
We then compare the results with a traditional MLE.
The computation times were measured using a typical commercial PC.
The result shows that the MLE has slightly better performance in terms of MSE;
however, the neural network also shows a reasonable result.
Meanwhile, the general numerical method of MLE requires many iterations and is time consuming.
However, a well-trained neural network computes an estimate very quickly, 
which is less than a hundredth of the time required by MLE.

\begin{center}
	\begin{tabular}{ccc}
		\hline
		& Neural network & MLE \\
		\hline
		MSE & 0.0513 & 0.0417\\
		Time (sec) & 0.0120 & 1.763 \\
		\hline
	\end{tabular}
\end{center}

%\begin{center}
%\begin{tabular}{ccc}
%	\hline
%	& neural network & MLE \\
%	\hline
%	MSE & 1.463 & 0.0148\\
%	Time & 0.012 & 1.763 \\
%	\hline
%\end{tabular}
%\end{center}

To understand the basic properties of the neural network estimator, we investigate its sampling distribution.
To examine the sampling distribution, we generate 100,000 paths of length 2,000 using the fixed parameter values of $\mu = 0.3$, $\alpha_{1} = 0.4$, $\alpha_{2} = 0.7$, $\beta = 1.5$. 
We then compare the obtained sampling distributions using the neural network and MLE methods.
Overall, MLE outperforms the neural network slightly.
Even so, the general performance of neural networks is also quite good.

\begin{center}
\begin{tabular}{cccccc}
	\hline
	Parameter & True & \multicolumn{2}{c}{Neural network} & \multicolumn{2}{c}{MLE} \\
	\hline
	& & Mean & S.D. & Mean & S.D. \\ 
	\hline
	$\mu$ & 0.3000 &  0.3151 & 0.0358 & 0.3036 & 0.0314 \\
	$\alpha_{1}$ & 0.4000 & 0.4429 & 0.0661 & 0.3988 & 0.0500\\
	$\alpha_{2}$ & 0.7000 & 0.5733 & 0.0697 & 0.7024 & 0.0608\\
	$\beta$ & 1.5000 & 1.5736 & 0.1364 & 1.5078 & 0.1145\\ 
	\hline
\end{tabular}
\end{center}

Specifically, the aforementioned example compares the performance of the numerical optimizer (used in MLE) and neural network.
Owing to the nature of simulation studies, the numerical optimizer has several advantages.
The outcome of a numerical optimizer is often influenced by its initial value.
In the example provided above, because the true value of the parameter is known, it was directly used as the initial value. 
This may have improved the performance compared to the result from a random initial value.
As finding a good initial value is sometimes challenging, 
a numerical optimizer may perform worse in real-world problems.

In addition, if the numerical optimizer exhibits unexpected behavior, such as failure to converge appropriately, 
human engagement may be necessary, such as adjustments to the initial value or retrying the procedure.
As the complexity of the model and level of noise present in the empirical data increase, 
the advantages of the numerical optimizer may decrease. 
In such cases, further research may be required to determine whether the numerical optimizer still outperforms the neural network.

\section{Empirical result}~\label{Sect:empirical}

The approach in the previous section can be directly applied to empirical data.
However, we need to consider whether robust estimation can be made in situations where the empirical data do not completely follow the Hawkes process. 
For example, in filtered high-frequency price process data, a subdue effect (where an event reduces intensity) can sometimes occur.
This can result in negative estimates for the parameter $\alpha$, which violates the definition of the Hawkes model.
To address this, a more complex model should be used;
however, as this falls outside the scope of this study, 
an alternative method is to use a softplus activation function $\log(1+\exp(x))$ for the last layer in the neural network. 
This approach is similar to constraint optimization by ensuring positive estimates.
Furthermore, instead of predicting $\beta$ directly, we trained and predicted $\beta-\alpha_1-\alpha_2$.
By using the softplus function, this method ensures that the branching ratio condition of the Hawkes model is met.

%In the empirical study, one of method is that
%we utilize a neural network trained on simulated data in the previous section 
%to estimate the parameters of a Hawkes process on real intra-day mid price data.
%Then compare the results obtained from the neural network with those obtained through the maximum likelihood estimation (MLE) method.
%Additionally, the intra-day Hawkes volatilities computed by Eq. based on estimates from the neural network and MLE also can be compared.

To further increase robustness, 
a combination of empirical data and its maximum likelihood estimates as training data can be used, 
rather than relying solely on simulation data. 
This approach accounts for the possibility of model mis-specification;
for instance, the observed data may not perfectly align with the Hawkes process. 
By incorporating the MLE into the training data, 
the neural network can better mimic the MLE of the Hawkes model. 
Thus, if the goal is to construct a neural network that closely approximates the MLE, even under the possibility of model mis-specification, this method can be effective.

The following section explains the step-by-step procedure.
We select segments of observed intraday data of inter-arrivals and movement types.
Each segment consists of a 2,000-time step.
These selected paths are used to fit the Hawkes model using MLE.
The resulting dataset is then used to train the neural network,
where the inter-arrivals and types of real data serve as feature variables and 
the maximum likelihood estimates are the target variables.

\begin{figure}[!hbt]
	\begin{subfigure}{.5\textwidth}
		\centering
		\includegraphics[width=0.94\textwidth]{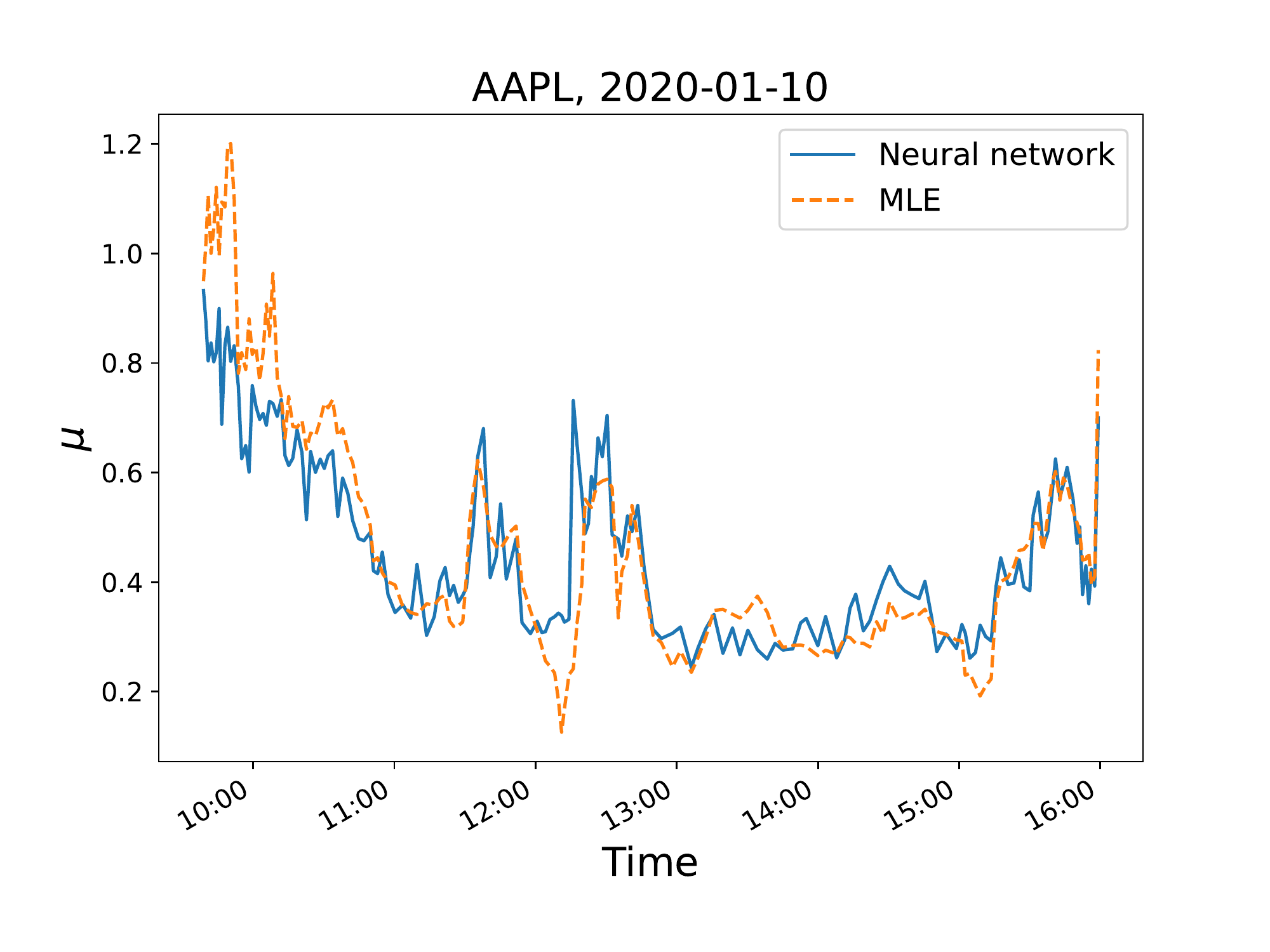}
		\caption{$\mu$}
		\label{fig:mu}
	\end{subfigure}
	\begin{subfigure}{.5\textwidth}
		\centering
		\includegraphics[width=0.94\textwidth]{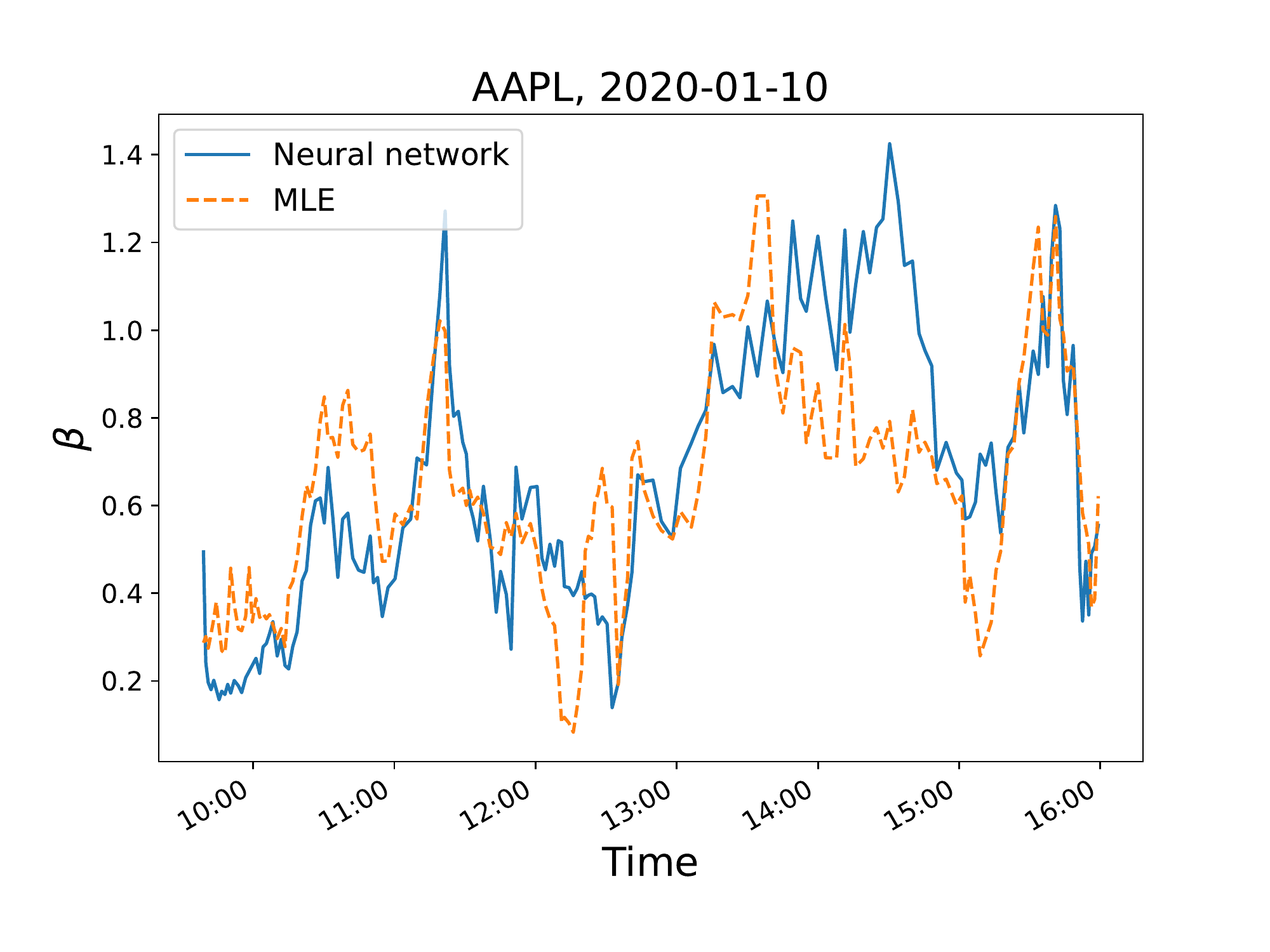}
		\caption{$\beta$}
		\label{fig:beta}
	\end{subfigure}
	
	\begin{subfigure}{.5\textwidth}
		\centering
		\includegraphics[width=0.94\textwidth]{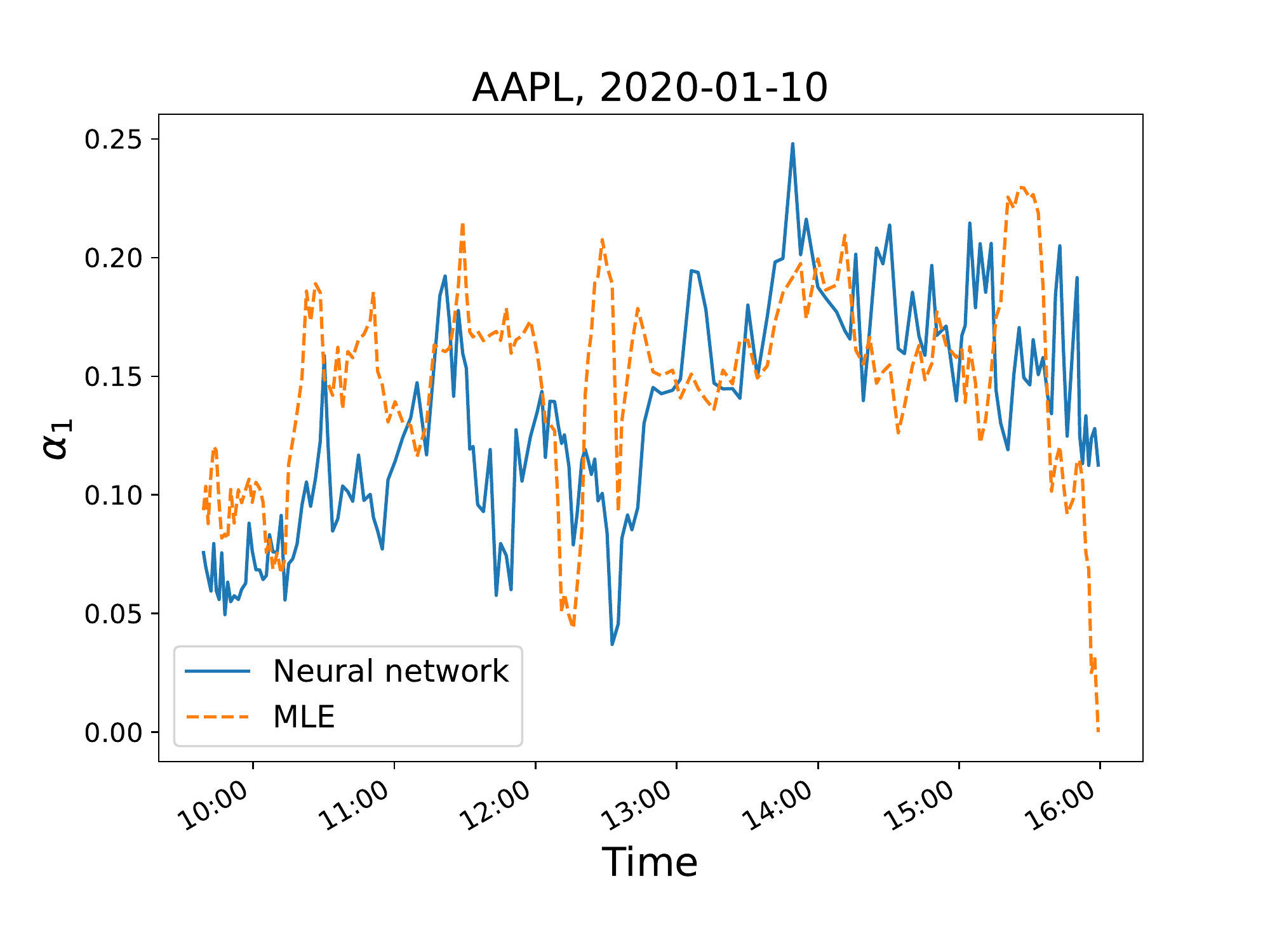}
		\caption{$\alpha_{1}$}
		\label{fig:alpha1}
	\end{subfigure}
	\begin{subfigure}{.5\textwidth}
		\centering
		\includegraphics[width=0.94\textwidth]{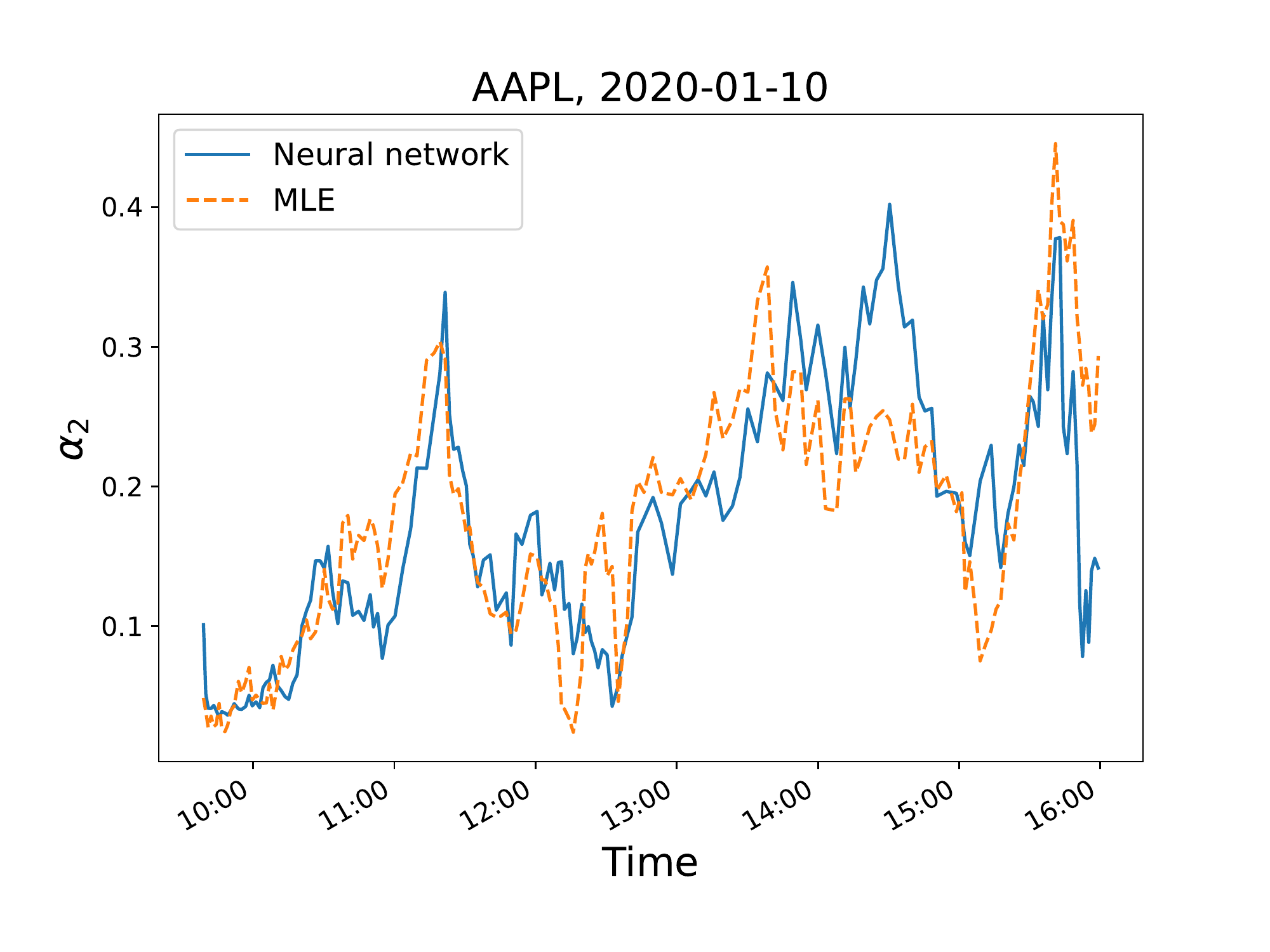}
		\caption{$\alpha_{2}$}
		\label{fig:alpha2}
	\end{subfigure}
	\caption{Intraday estimates for the NBBO of AAPL using MLE and neural network}
	\label{Fig:estimates}
\end{figure}

Figure~\ref{Fig:estimates} illustrates the intraday dynamics of estimates of the Hawkes model on a specific date. 
The data used for this illustration are out-of-sample that are not used for training.
Specifically, it was estimated using segments of data corresponding to every 2,000-time step. 
This corresponds to a time horizon of approximately 10-20 minutes.
To create a more continuous graph, the time windows for estimation were moved forward slowly with sufficient overlap.
The neural network shows very consistent results with MLE.
%Overall, immediately after the market opened, it was observed that the estimates of $\alpha$s and $\beta$ in both MLE and neural network are less significant than other periods, which makes the price process less likely to have the Hawkes property.

\begin{figure}[!hbt]
	\begin{subfigure}{.5\textwidth}
		\centering
		\includegraphics[width=0.94\textwidth]{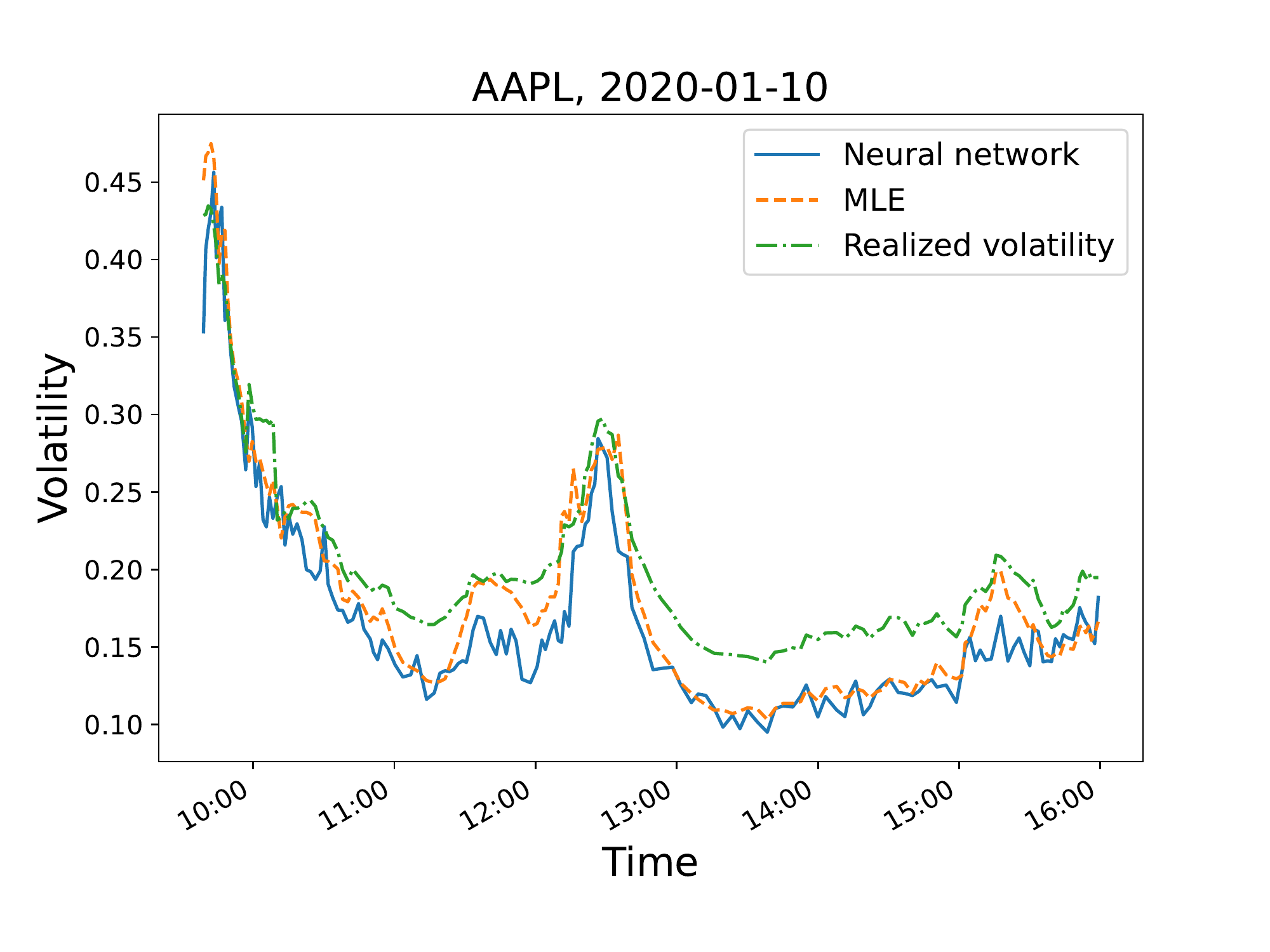}
	\end{subfigure}
	\begin{subfigure}{.5\textwidth}
		\centering
		\includegraphics[width=0.94\textwidth]{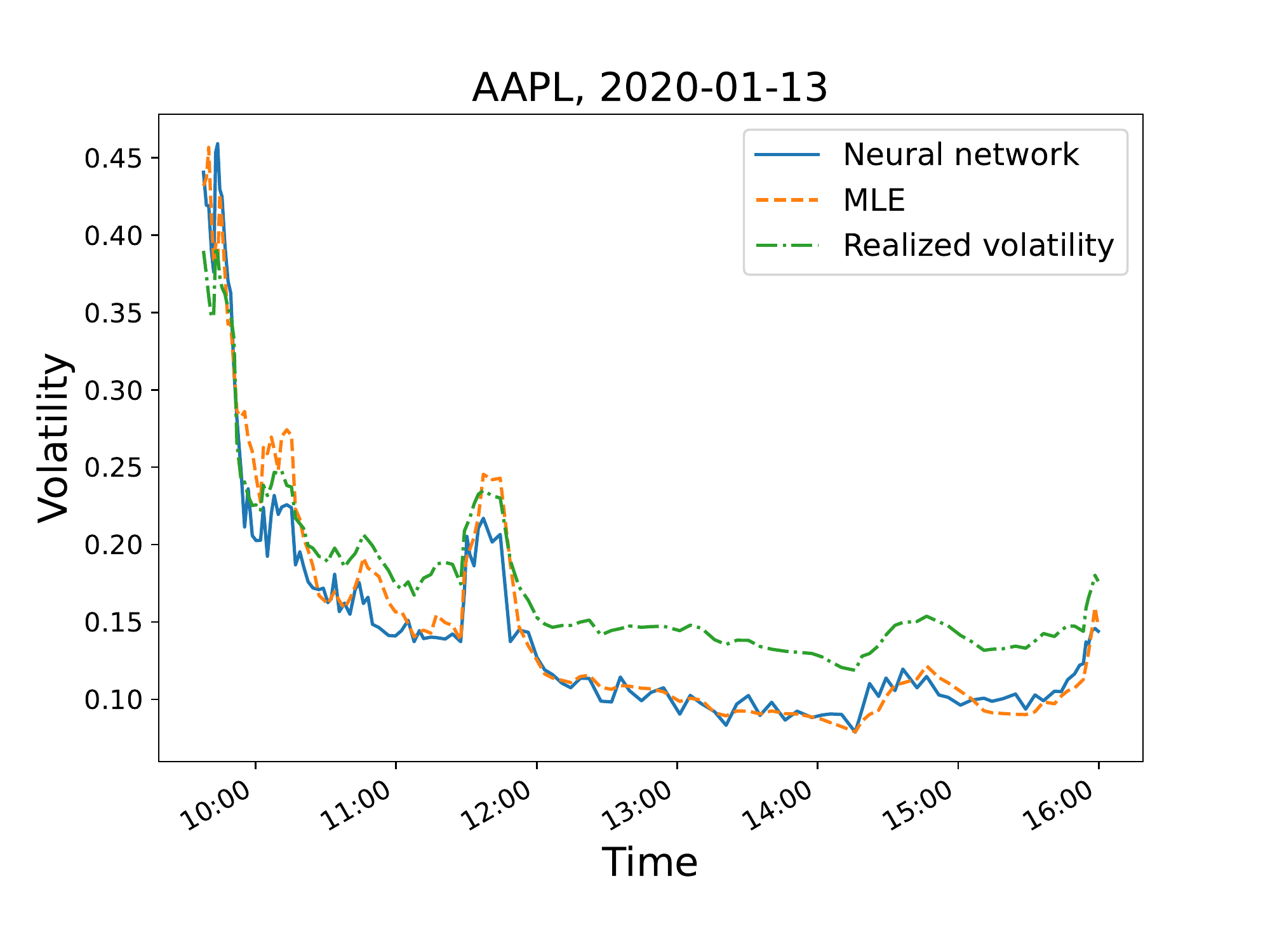}
	\end{subfigure}
	\caption{Intraday annualized volatilities for the NBBO of AAPL using MLE and neural network}
	\label{Fig:vol}
\end{figure}

Figure~\ref{Fig:vol} presents the instantaneous intraday annualized Hawkes volatility calculated using MLE, a neural network, and nonparametric realized volatility as a benchmark by \cite{ABDL}. 
The realized volatility is calculated using the observed values at 1-second intervals for the price process of the period. 
All three measures have a similar trend throughout the day.
Although it is not possible to present all the results examined, in some cases, 
MLE showed unstable dynamics of volatility.
This is likely due to the fact that the 2,000-time length used for estimation is a relatively small sample size to estimate the parameters of the Hawkes model.

\section{Conclusion}~\label{Sect:concl}
This study shows that a neural network can accurately estimate time series parameters, 
with an accuracy similar to MLE and much faster computation. 
While the example used here is for calculating Hawkes volatility, 
but our proposed method can be applied to various fields. 
It can be particularly useful in cases where the model is complex and traditional estimation procedures are challenging, 
such as modeling entire limit order book. 
Further research in this area is expected to be ongoing and diverse.

\section*{Acknowledgements}
\noindent
This work has supported by the National Research Foundation of Korea(NRF) grant funded by the Korea government(MSIT)(No. NRF-2021R1C1C1007692).

%\bibliography{Bib}
%\bibliographystyle{chicago}

\end{document}